# Tuning photoluminescence response by electric field in the lead-free piezoelectric $Na_{1/2}Bi_{1/2}TiO_3$-$BaTiO_3$

Deepak Kumar Khatua, Abhijeet Kalaskar and Rajeev Ranjan*

[1]Department of Materials Engineering, Indian Institute of Science Bangalore-560012, India

## Abstract

We show that an electrically soft ferroelectric host can be used to tune the photoluminescence (PL) response of rare-earth emitter ions by external electric field. The proof of this concept is demonstrated by changing the PL response of $Eu^{+3}$ ion by electric field on a model system Eu-doped $0.94Na_{1/2}Bi_{1/2}TiO_3$-$0.06BaTiO_3$. We also show that new channels of radiative transitions, forbidden otherwise, open up due to positional disorder in the system, which can as well be tuned by electric field.

* ranjanrajeeb@gmail.com

Photoluminescence (PL) property of lanthanide ions is a subject of continuing interest because of its enormous applications in lighting industry, Laser, fiber optics telecommunications, biological assaying and medical imaging [1, 2]. The most common strategy prevalent among experimentalists is to investigate the PL response/mechanisms by placing the luminescent centers in different kinds of host matrices or ligand environments. Among others, crystallographic symmetry of the host crystal is one of the important factors that influences PL emission in solids [3]. As such, temperature, pressure, and compositional modifications which can induce crystallographic transformations can, as well, affect the PL response. In general, pressure induced phase transformations in inorganic materials occur at Giga Pascal pressure ranges which is not easily achievable under ordinary conditions. From a device point of view, it is always desirable to tune/control properties around the ambient temperature and pressure conditions. In principle, it can be argued that if, apart from temperature and pressure, other control parameters are available to change the crystal structure, this parameter can as well tune different types of structure-sensitive properties, including PL. Such a scenario does exist in electrically and elastically soft ferroelectrics in the proximity of a ferroelectric instability. The electromechanical softness of the lattice, in conjunction with the strong coupling of polarization with crystal structure, makes such systems amenable to field induced structural transformation [4-13]. The consequent change in the crystal field parameters is expected to affect the PL response of doped emitter ions, such as the rare earth ions, if they are embedded in such a ferroelectric host. Apart from serving as probes to understand the local structural changes due to external influences such as temperature, pressure, electric field, etc., rare-earth doping of soft ferroelectrics can form a new class of multifunctional materials, which may be termed as "luminescent-piezoelectric", based on which new applications can be envisaged. With this as the motivation, and to demonstrate the proof of this concept, we have studied the electric field effect on the PL response of $Eu^{+3}$ ion by doping them in a lead-free piezoelectric $0.94Na_{1/2}Bi_{1/2}TiO_3$-$0.06BaTiO_3$ (NBT-6BT). NBT-6BT is one of the most extensively investigated lead-free piezoelectric in the recent past [14-16]. The first report by Takenaka et al attributed the anomalous piezo-properties in NBT-6BT to a rhombohedral-tetragonal instability and coexistence of ferroelectric phases [14]. Subsequent studies reported that the critical composition exhibits cubic structure [17], and the hetrophase ferroelectric state appears only after application of strong electric field, i.e. poling [13, 16, 18]. In view of the drastic effect of electric field on this system, we chose NBT-6BT as a model host for investigating the effect of electric field on the PL response. The ion chosen for the PL study was $Eu^{+3}$, as this is known to be an efficient emitter in the red region. For decades, Eu-doped $Y_2O_3$ has been used as a red phosphor in fluorescent lamps [19]. We carried out a systematic study of the effect of electric field on the PL response of the $Eu^{+3}$ ion, in conjunction with structural, dielectric and ferroelectric response at room temperature, as well as close to its nonergodic-ergodic crossover regime, on a dilute Eu-doped NBT-6BT, i.e $0.94Na_{0.5}Bi_{0.495}Eu_{0.005}TiO_3$–$0.06BaTiO_3$ (NBT-6BT:Eu). Our results prove the validity of the conjecture that electric fields can tune the PL response of the doped rare earth ions if they are hosted in an electrically soft piezoelectric. In the process, we also discovered that PL of $Eu^{+3}$ acts as a tool to probe the degree of positional disorder in this system.

Specimens of $0.94Na_{0.5}Bi_{0.495}Eu_{0.005}TiO_3$ –$0.06BaTiO_3$ (NBT-6BT:Eu) were prepared by conventional ceramic synthesis route. The details of the experimental methods are provided in the Supplemental information S1. Both NBT-6BT, and its Eu-doped variant, NBT-6BT:Eu, exhibited identical longitudinal direct-piezoelectric coefficient ($d_{33}$), ~127 pC/N and 126 pC/N, respectively, confirming that the dilute concentration of Eu did not alter piezoelectric properties of the host system. At the same time, this specimen gave a very good $Eu^{+3}$ PL signal (Fig. 1). It is also important to note that in the wavelength range of interest with regard to the $Eu^{+3}$ PL (570 – 720 nm), there is no PL signal from the host matrix (Supplemental S2). Following a comprehensive review on $Eu^{+3}$ photoluminescence [3], the emission lines in the different wavelength ranges are categorized as $^5D_0 \rightarrow {}^7F_0$ (570-585 nm), $^5D_0 \rightarrow {}^7F_1$ (585-600 nm), $^5D_0 \rightarrow {}^7F_2$ (610-630 nm), $^5D_0 \rightarrow {}^7F_3$ (640-660 nm), and $^5D_0 \rightarrow {}^7F_4$ (680-710 nm). The multiplets associated with these transitions arise due to crystal field splitting. The degeneracy of the different $^7F_J$ levels is completely lifted for $Eu^{+3}$ ion an orthorhombic or lower symmetry, leading to the splitting of the $^7F_J$ level into 2J+1 components.

Next we discuss the effect of electric field on the $Eu^{+3}$ PL spectrum of NBT-6BT:Eu. For a better insight with regard to the mechanism associated with the field effect on PL, we also investigated the effect of electric field on the $Eu^{+3}$ PL in $Na_{1/2}Bi_{1/2}TiO_3$ (NBT), Fig. 2. The concentration of Eu in NBT was kept the same as in NBT-6BT:Eu, i.e. $Na_{0.5}Bi_{0.495}Eu_{0.005}TiO_3$ (NBT:Eu). As a general observation, it can be seen that the PL spectrum of NBT-6BT:Eu tends to resemble the PL spectrum of NBT: Eu after poling. For example, while the $^5D_0 \rightarrow {}^7F_2$ Stark band of unpoled NBT-6BT:Eu shows two distinct Stark lines at 615 nm and 625 nm, the $^5D_0 \rightarrow$

$^7F_2$ Stark band of the poled specimen exhibits an additional emission line close to 621 nm, making this band appear as a triplet, which is similar to the triplet nature of this Stark band of NBT:Eu. Vanishing of a Stark line at 587 nm in the $^5D_0 \rightarrow ^7F_1$ Stark band also occurs after poling NBT-6BT:Eu. Another interesting feature to note is that poling decreased the relative intensity of the strongest line at 593 nm in the $^5D_0 \rightarrow ^7F_1$ Stark band decreased from 78 % to 68 %. This decrease is halfway the intensity between unpoled NBT-6BT:Eu (78 %) and NBT:Eu (58%), and can be rationalized by the fact that nearly half the volume fraction of the cubic-phase of the unpoled NBT-6BT:Eu (Fig. 3a) transforms to rhombohedral after poling (Fig. 3b, Supplemental S3). Poled NBT:Eu, on the other hand, exhibits a pure rhombohedral strucrture (Fig. 3c, Supplemental S4). Experiments in the past have shown that the $^5D_0 \rightarrow ^7F_1$ magnetic dipole transition dominates the $Eu^{+3}$ PL spectrum in cubic host crystals such as $Ba_2GdNbO_6$:Eu [20], $SrTiO_3$: Eu [21] $SrSnO_3$ : Eu [22]. The comparatively large intensity (78 %) of the $^5D_0 \rightarrow ^7F_1$ line in unpoled NBT-6BT:Eu can be attributed to the cubic average structure. However, the fact that the $^5D_0 \rightarrow ^7F_1$ Stark profile shows superposition of more than two lines confirms that the local ligand environment around Eu has non-cubic symmetry [3]. A possibility of long period structural modulation has earlier been suggested by Garg et al [18].

The effect of electric field on the $^5D_0 \rightarrow ^7F_0$ Stark line near 580 nm needs special mention. Irrespective of the crystallographic symmetry, this line is expected to be a singlet. However, as can be seen from the inset of Fig. 1a, this line is asymmetrically broadened in unpoled NBT-6BT:Eu suggesting two closely spaced lines of nearly equal intensity (marked with an arrow). Earlier, extensive broadening of the $^5D_0 \rightarrow ^7F_0$ line have been reported in Eu doped calcium diborate, silicate, germinate and phosphate glasses due to a large number of sites of same local symmetry ($C_s$) but with slightly varying crystal field strengths [23, 24]. In our case, the slightly different crystal field strengths can arise from the different local chemical environment around $Eu^{+3}$ formed by different configurations of Na and Bi and Ba [25, 26], along with their respective local polar displacements [27]. There is a remarkable decrease in the broadening of this peak after poling NBT-6BT:Eu as shown in the inset of Fig. 1b. As shown in the inset of Fig. 3a, although with a lesser intensity, additional shoulder in the $^5D_0 \rightarrow ^7F_0$ profile is evident in the PL of NBT:Eu. This shoulder disappears in the poled NBT:Eu. In conformity with the recent reports on pure NBT [28, 29], the XRD pattern of the unpoled and poled NBT:Eu could fit with a monoclinic (*Cc*) and rhombohedral (R3c) structural models, respectively (supplemental S4). The fact that the *Cc* distortion completely vanishes after poling suggests that *Cc* is not likely to be a thermodynamic phase of NBT at room temperature. Subsequent studies associated the monoclinic (*Cc*) like average distortion to the presence of R3c incompatible short ranged in-phase ($a^0a^0c^+$) octahedral tilts [30, 31], and displacements of Na/Bi cation away from the [111] polar direction of the R3c phase [27]. In the process of establishing a long range ferroelectric order, strong electric field "wipes away" the positional disorder and the associated structural heterogeneity on the mesoscopic scale, leaving the system in a pure rhombohedral ferroeletric phase [32]. According to this analysis, the *Cc* phase in NBT is a manifestation of positional disorder in the otherwise polar rhombohedral matrix. In view of this, the additional shoulder in the $^5D_0 \rightarrow {}^7F_0$ line of unpoled NBT:Eu and its disappearance after poling, prove that positional disorder is the primary factor for the additional shoulder in the $^5D_0 \rightarrow {}^7F_0$ line. That is, positional disorder provides additional channels of radiative transition, forbidden otherwise, and that these channels can be made to vanish on application of electric field. We would like to emphasize that the PL spectra were highly reproducible, and that the changes in the relative intensities discussed above are beyond fluctuations in counting statistics. As a proof of this, we separately recorded the PL of an unpoled NBT-6BT:Eu and a de-poled NBT-6BT:Eu. The two normalized PL spectra exactly superposed (S5 of the supplemental formation), thereby confirming that changes in relative intensities are unambiguous, and that the interpretations derived from them are meaningful.

Although the significant broadening of the $^5D_0 \rightarrow {}^7F_0$ transition in unpoled NBT-BT6-Eu is associated with the local positional disorder, the Bragg profiles of NBT-BT6-Eu does not exhibit noticeable broadening of the peaks. For example the FWHM of (111) and (200) peaks are 0.06 $^o$ and 0.08 $^o$, respectively, which is comparable to the instrumental resolution of the diffractometer. This implies that the structural disorder does not introduce significant random lattice strain in NBT-BT6-Eu. A close inspection of the Bragg profiles, however, reveals gradually decaying tails suggesting considerable diffuse scattering around the fundamental cubic reflections. Such diffuse scattering originate from short ranged correlation in atomic displacement, and have been investigated extensively in $Na_{1/2}Bi_{1/2}TiO_3$ [33, 34] and $Pb(Mg_{1/3}Nb_{2/3})O_3$ [35]. The short ranged correlation in the atomic displacements makes these systems behave a relaxor ferroelectric with pronounced dielectric relaxation. Since we have identified the additional shoulder in the $^5D_0 \rightarrow {}^7F_0$ Stark line of NBT:Eu with positional disorder in the system, the considerably enhanced intensity of this shoulder in unpoled NBT-6BT:Eu suggests that the average cubic structure is associated with a relatively greater degree of positional disorder as compared to that in NBT:Eu. To verify this, we compared the dielectric dispersion of the two systems in their poled and unpoled states at room temperature, Fig. 4. It is evident from this figure that the relative permittivity of unpoled NBT-6BT:Eu decreases from 2000 to 1400, i.e. by 600, as the frequency increases from 100 Hz to 1 MHz. Poling led to a significant decrease in the relative permittivity as well as the frequency dispersion. For example, poling decreases the relative permittivity from 2000 to 980 at 100 Hz.The significant decrease in the slope of the frequency-permittivity plot after poling is a manifestation of reduced dielectric relaxation as compared to that in the unpoled state. It may, however, be noted that the relative permittivity of poled NBT-6BT:Eu is still higher than the relative permittivities of NBT:Eu which can be attributed to the fact that the poled NBT-6BT:Eu still has ~45 % cubic (relaxor) phase (Supplemental S3).

The results presented above pertain to irreversible change in the structure and the PL response after application of electric field at room temperature. We also carried out field dependent PL study close to the nonergodic – ergodic crossover temperature of the NBT-6BT:Eu. In the present case, this temperature also happens to be the depolarization temperature of the system. Temperature dependent dielectric measurement of a poled specimen of NBT-6BT:Eu revealed the depolarization temperature to be ~85 $^{o}$C (See supplemental information S6). The variation in the PL spectra on application of electric field at 80 $^{o}$C is shown in Fig. 5. A distinct change in the PL spectrum can be seen when the specimen is subjected to a field of ~ 35 kV/cm at this temperature. For example, the relative intensity of the strongest line in the $^{7}F_{1}$ Stark band decrease from 82 % to 72 %. The Field also suppresses the intensity of the line at 587 nm in this Stark band. Similarly, a new line seems to develop at 622 nm. These trends are identical to what was observed at room temperature after poling NBT-6BT:Eu. The important difference, however, is that the spectrum recovered when the field was switched off at 80 $^{o}$C. This experiment suggests that the PL can be reversibly tuned if the system is in the proximity of its ergodic-nonergodic crossover regime. If, by suitable compositional modification, the ergodic-nonergodic crossover/depolarization temperature is brought close to room temperature, it should be possible to tune the PL response by electric field in a reversible manner at room temperature.

Although in this Letter we have demonstrated the feasibility of the tuning the PL response of $Eu^{+3}$ by hosting it in an NBT-based lead-free relaxor ferroelectric, we expect a similar tuning to be possible if $Eu^{+3}$ is hosted in another relaxor ferroelectric such as $Pb(Mg_{1/3}Nb_{2/3})O_3$ and La-modified $Pb(Zr, Ti)O_3$ [36]. The difference in the chemical environment around $Eu^{+3}$ in different relaxor ferroelectric systems will however result in different crystal field strengths at the $Eu^{+3}$ site, giving rise to slight differences in the PL spectra. While the details of structural changes induced by electric field is likely to differ from system to system, the generic result of fundamental significance is that the PL properties of emitter ions can be tuned by electric field if they are doped in a host which is electrically soft. This idea is equally relevant to rotator ferroelectrics such as $Pb(Ti, Zr)O_3$, $Pb(Mg_{1/3}Nb_{2/3})O_3$-$PbTiO_3$, and $Pb(Zn_{1/3}Nb_{2/3})O_3$-$PbTiO_3$ which undergo polarization switching and field induced structural transformation [4-8]. In the recent past field induced phase transformation has been demonstrated in several systems such as soft-PZT [9, 36], $BiScO_3$-$PbTiO_3$ [10, 11], $BaTiO_3$ based systems [12, 37-40], Li-modified $Na_{0.5}K_{0.5}NbO_3$ [41]. Remarkable structural changes by field is also possible in antiferroelectrics systems in the vicinity of antiferroelectrics-ferroelectric instability, and such systems are also potential candidates for electric-field tuning of PL.

In conclusion, we have shown that by incorporating $Eu^{+3}$ in the host lattice of a lead-free relaxor ferroelectric, $0.94Na_{1/2}Bi_{1/2}TiO_3$-$0.06BaTiO_3$, the photoluminescence PL response of the guest ion can be tuned by an electric field. We use this experimental demonstration to argue that this phenomenon can, in principle, be seen in all electrically soft ferroelectrics or antiferroelectrics due to their propensity for field induced structural transformation. The change in the PL response is associated with the field induced changes in the local symmetry and crystal field strength at the guest ion site. We also show that in case the host is a relaxor ferroelectric, the PL response of the guest emitter ion can act as a local probe to determine the relative degree of positional polar disorder as a function of composition/temperature/electric field, as positional disorder opens new channels for radiative transitions. We hope that this work will stimulate further studies on similar lines.

**Acknowledgments:** RR thanks the Science and Engineering Board (SERB) of the Department of Science and Technology, Govt. of India for financial support (Grant No. SERB/F/5046/2013-14).

**References**

[1] J. C. G. Bunzli, and C. Piguet, Chem. Soc. Rev. **34**, 1048 (2005)

[2] J. C. G. Bunzli, Coord. Chem. Rev. **293–294**, 19 (2015)

[3] K. Binnemans, Coord. Chem. Rev. **295**, 1 (2015).

[4] H. Fu and R.E. Cohen, Nature **403**, 281 (2000).

[5] D. Damjanovic J Am Ceram Soc **88**, 2663 (2005)

[6] D. Damjanovic, Appl. Phys. Lett. **97**, 062906 (2010)

[7] L. Bellacihe, A. Garcia, and D. Vanderbilt, Phys. Rev. B **64**, 060103 (2001)

[8] D. Vanderbilt and M. H. Cohen, Phys. Rev. B **63**, 094108 (2001)

[9] M. Hinterstein, J. Rouquette, J. Haines, Ph. Papet, M. Knapp, J. Glaum and H. Fuess, Phys. Rev. Lett. **107**, 077602 (2011).

[10] Lalitha K V, A. N. Fitch and R. Ranjan, Phys. Rev. B **87**, 064106 (2013)

[11] Lalitha K V, A. K. Kalyani and R. Ranjan, Phys. Rev. B **90** 224107 (2014)

[12] A. K. Kalyani, H. Krishnan, A. Sen, A. Senyshyn, and R. Ranjan, Phys. Rev. B **91**, 024101 (2015)

[13] R. Garg, B. N. Rao, A. senyshyn, P. S. R. Krishna, and R. Ranjan, Phys. Rev. B **88**, 014103 (2013)

[14] T. Takenaka, K. Maruyama, and K. Sakata, Jpn. J. Appl. Phys. **30**, 2236 (1991).

[15] J. Rödel, W. Jo, K. T. P. Seifert, E.-M. Anton, T. Granzow, and D. Damjanovic, J. Am. Ceram. Soc. **92**, 1153 (2009).

[16] W. Jo, J.E. Daniels, J.L. Jones, X. Tan, P.A. Thomas, D. Damjanovic, and J. Rödel, J. Appl. Phys. **109**, 014110 (2011).

[17] R. Ranjan and A. Dviwedi, Solid State Commun. **135**, 394 (2005).

[18] R. Garg, A. Senyshyn and R. Ranjan, J. Appl. Phys. **114**, 234102 (2013)

[19] N. C. Chang, J. Appl. Phys. **34**, 3500 (1963)

[20] G. Blasse, A. Bril, W.C. Nieuwpoort, J. Phys. Chem. Solids **27**, 1587 (1966).

[21] F. Fujishiro, T. Arakawa, T. Hashimoto, Mater. Lett. **65**, 1819 (2011)

[22] S. Basu, D. K. Patel, J. Nuwad, V. Sudarsan, S. N. Jha, D. Bhattacharya, R. K. Vatsa, S. K. Kulshreshtha, Chem. Phys. Lett. **561**, 82 (2013).

[23] V. Lavin, U. R. Rodriguez-Mendoza, I. R. Martin, V. D. Rodriguez, J. Non-Cryst. Solids **319**, 200 (2003).

[24] R. Reisfeld, R. A. Velapoldi, L. Boehm, M. Ish-Shalom, J. Phys. Chem. **75**, 3980 (1971).

[25] M. Gröting, I. Kornev, B. Dkhil, and K. Albe, Phys. Rev. B **86**, 134118 (2012)

[26] M. Gröting and K. Albe, Phys. Rev. B **89**, 054105 (2014)

[27] J. Kreisel, A.M. Glazer, P. Bouvier, and G. Lucazeau, Phys. Rev. B **63**, 174106 (2001)

[28] E. Aksel, J.S. Forrester, J.L. Jones, P.A. Thomas, K. Page, and M.R. Suchomel, Appl. Phys. Lett. **98**, 152901 (2011).

[29] B.N. Rao and R. Ranjan, Phys. Rev. B **86**, 134103 (2012).

[30] B. N. Rao, R. Datta, S. S. Chandrashekaran, D. K. Mishra, V. Sathe, A. Senyshyn and R. Ranjan, Phys. Rev. B **88** 224103 (2013)

[31] I. Levin and I. M. Reaney, Adv. Funct. Mater. **22**, 3445 (2012).

[32] B. N. Rao, L. Olivi, V. Sathe, and R. Ranjan, Phys. Rev. B **93**, 024106 (2016)

[33] J. Kreisel, P. Bouvier, M. Maglione, B. Dkhil, A. Simon, Phys. Rev. B. **69**, 092104 (2004)

[34] M. Matsura, H. Ida, K. Horita, K. Ohwada, Y. Noguchi and M. Miyayama, Phys. Rev B **87**, 064109 (2013)

[35] S. B. Vakhrushev, A. A. Naberezhnov, N. M. Okuneva, and B. N. Savenko, Fiz. Tverd. Tela (St. Petersberg) **37**, 3621 (1995) [Sov. Phys. Solid State **37**, 1993 (1995)]

[36] G. A. Samara, J. Phys.: Condens. Matter **15**, R367 (2003).

[37] A. K. Kalyani, Lalitha K V, A. R. James, A. N. Fitch and R. Ranjan, J. Phys: Condens. Matter **27**, 072201 (2015)

[38] A. K. Kalyani and R. Ranjan, J. Phys.: Condens. Matter **25**, 362203 (2013).

[39] A. K. Kalyani, D. K. Khatua, B. Loukya, R. Datta, A. N. Fitch, A. Senyshyn and R. Ranjan, Phys. Rev. B. **91**, 104104 (2015).

[40] K. Brajesh, K. Tanwar, M. Abebe and R. Ranjan, Phys. Rev. B **92**, 224112 (2015).

[41] T. Iamsasri, G. Tutuncu, C. Uthaisar, S. Wongsaenmai, S. Pojprapai, and J. L. Jones, Appl. Phys. Lett. **117**, 024101 (2015).

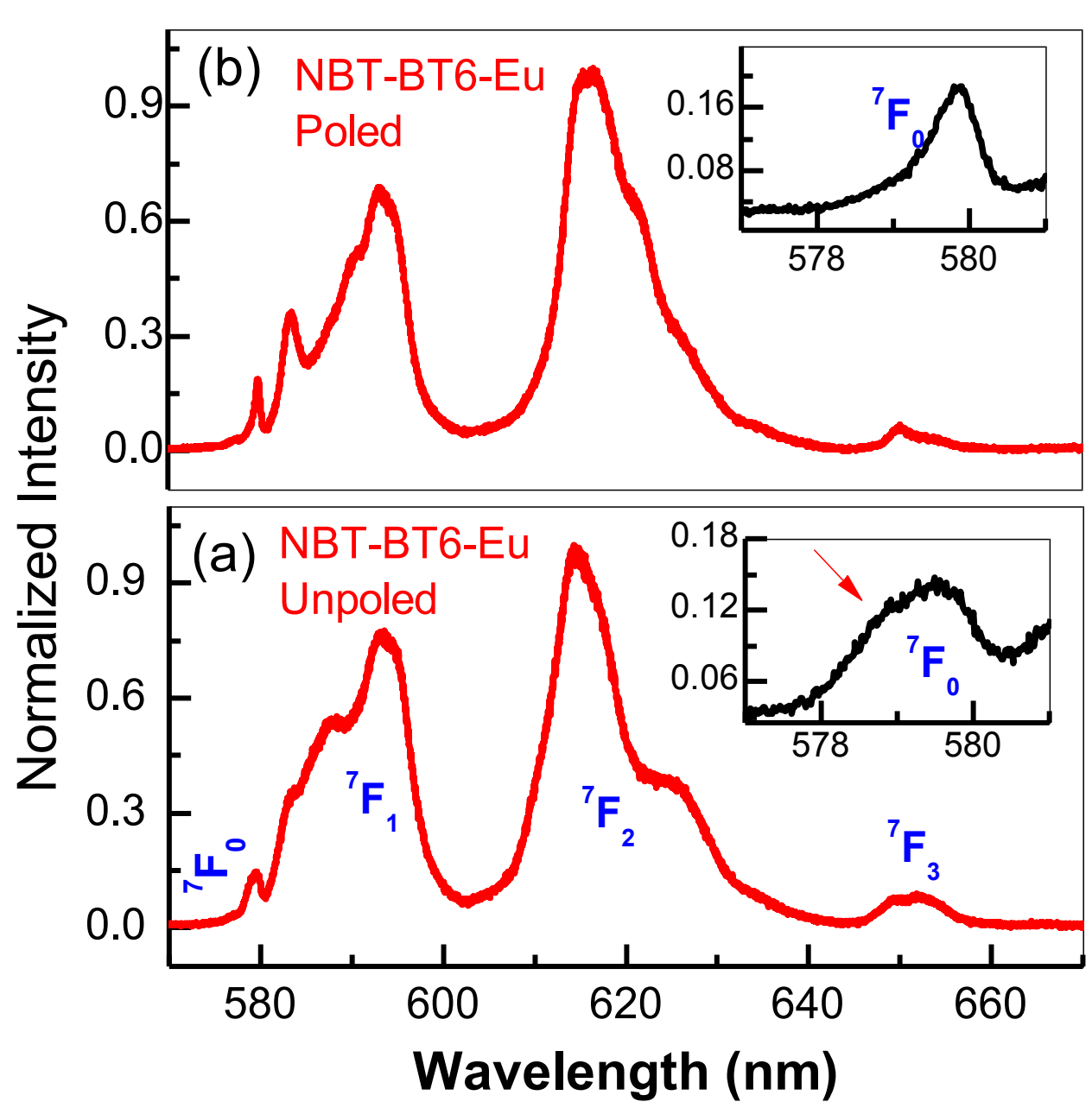

Fig. 1 Photoluminescence spectra of NBT-6BT: Eu poled and unpoled specimens. The insets highlight the magnified view of the $^5D_0 \rightarrow {}^7F_0$ Stark line.

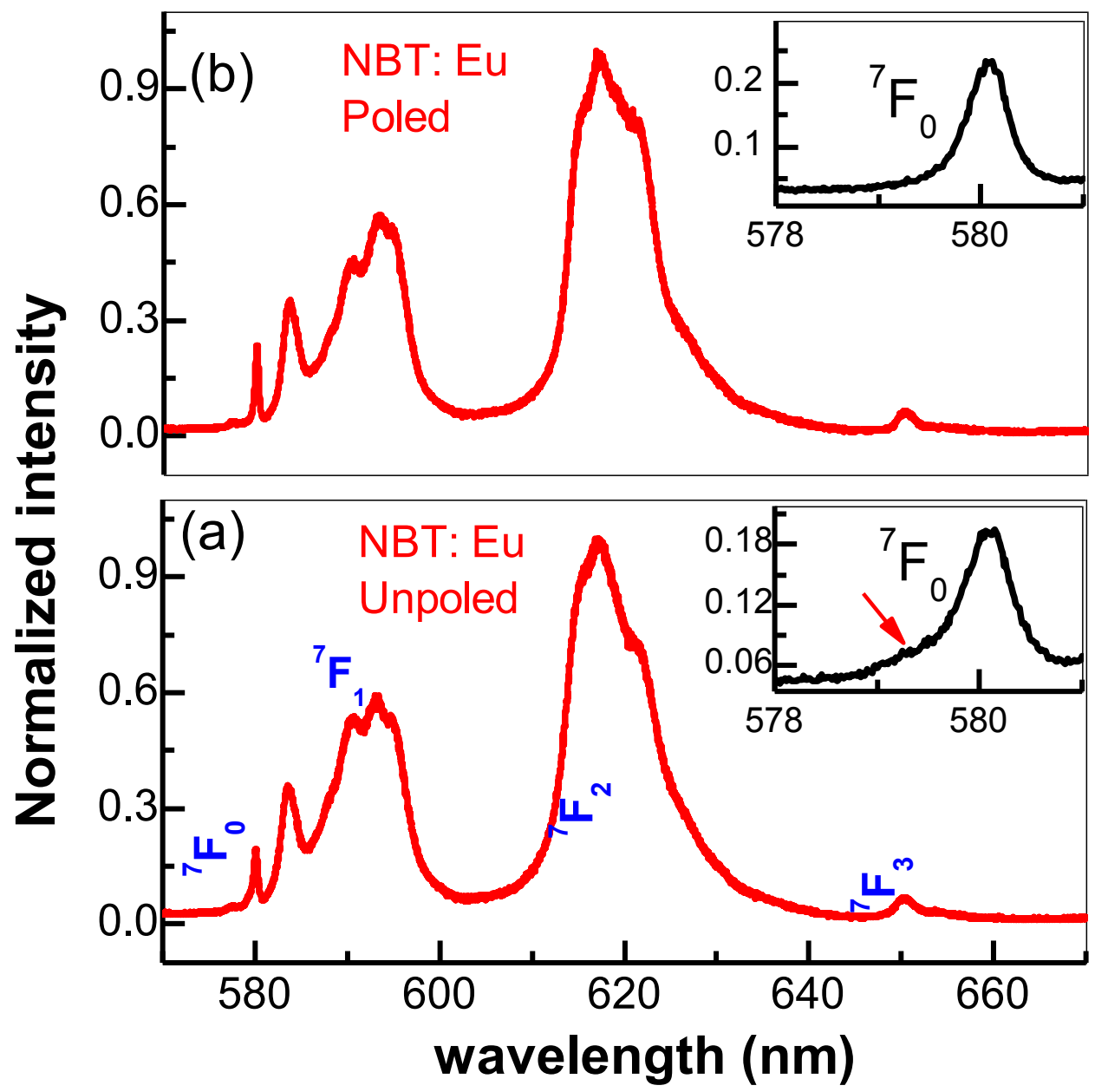

Fig.2 Photoluminescence spectra of NBT: Eu poled and unpoled specimens. The insets highlight the magnified view of the $^5D_0 \rightarrow {}^7F_0$ Stark line.

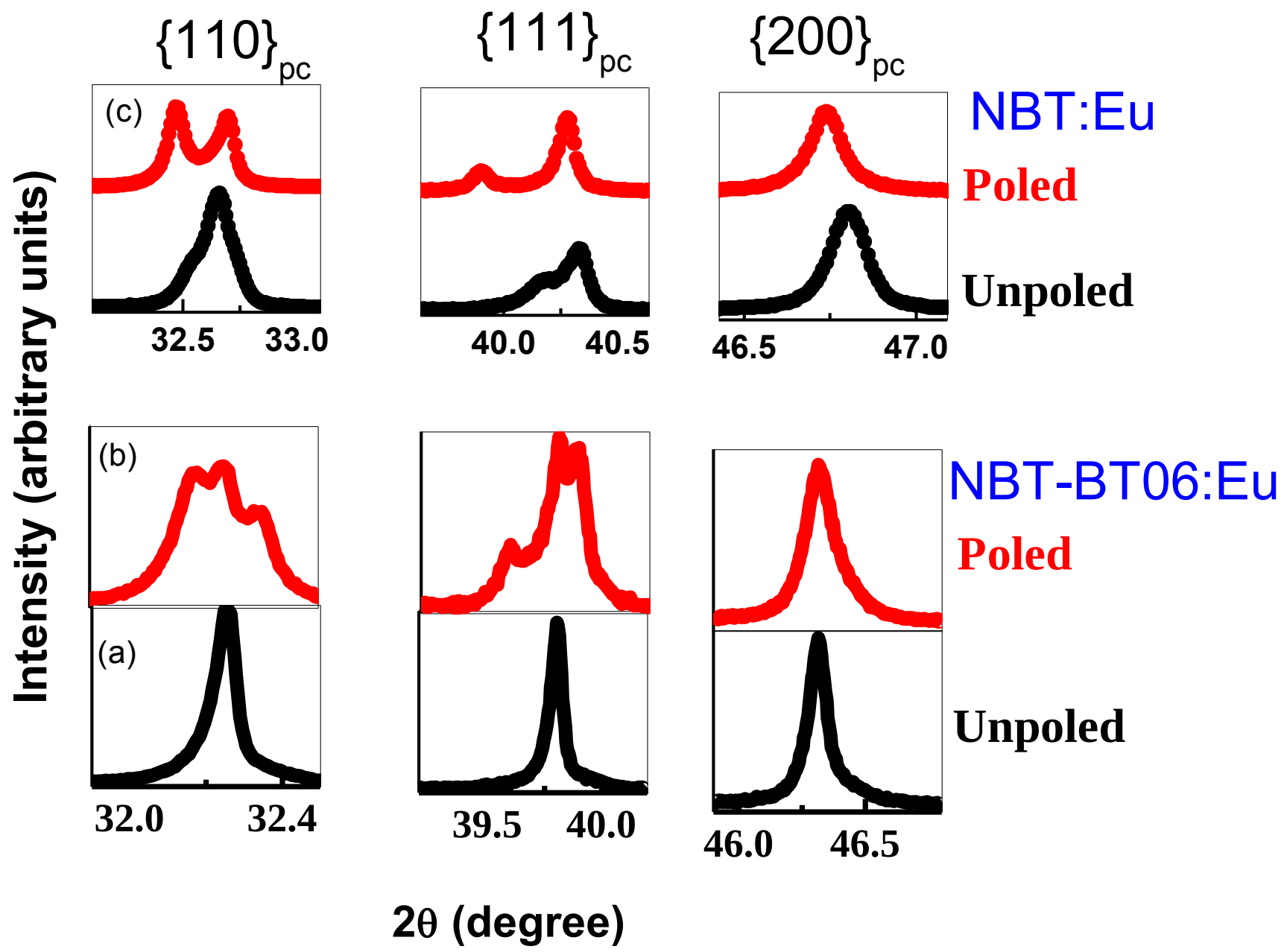

Fig. 3Three characteristic Bragg profiles (in the pseudocubic notation their indices have been written as $\{110\}_{pc}$, $\{111\}_{pc}$ and $\{200\}_{pc}$) of NBT:Eu and NBT-6BT:Eu for unpoled and and poled specimens.

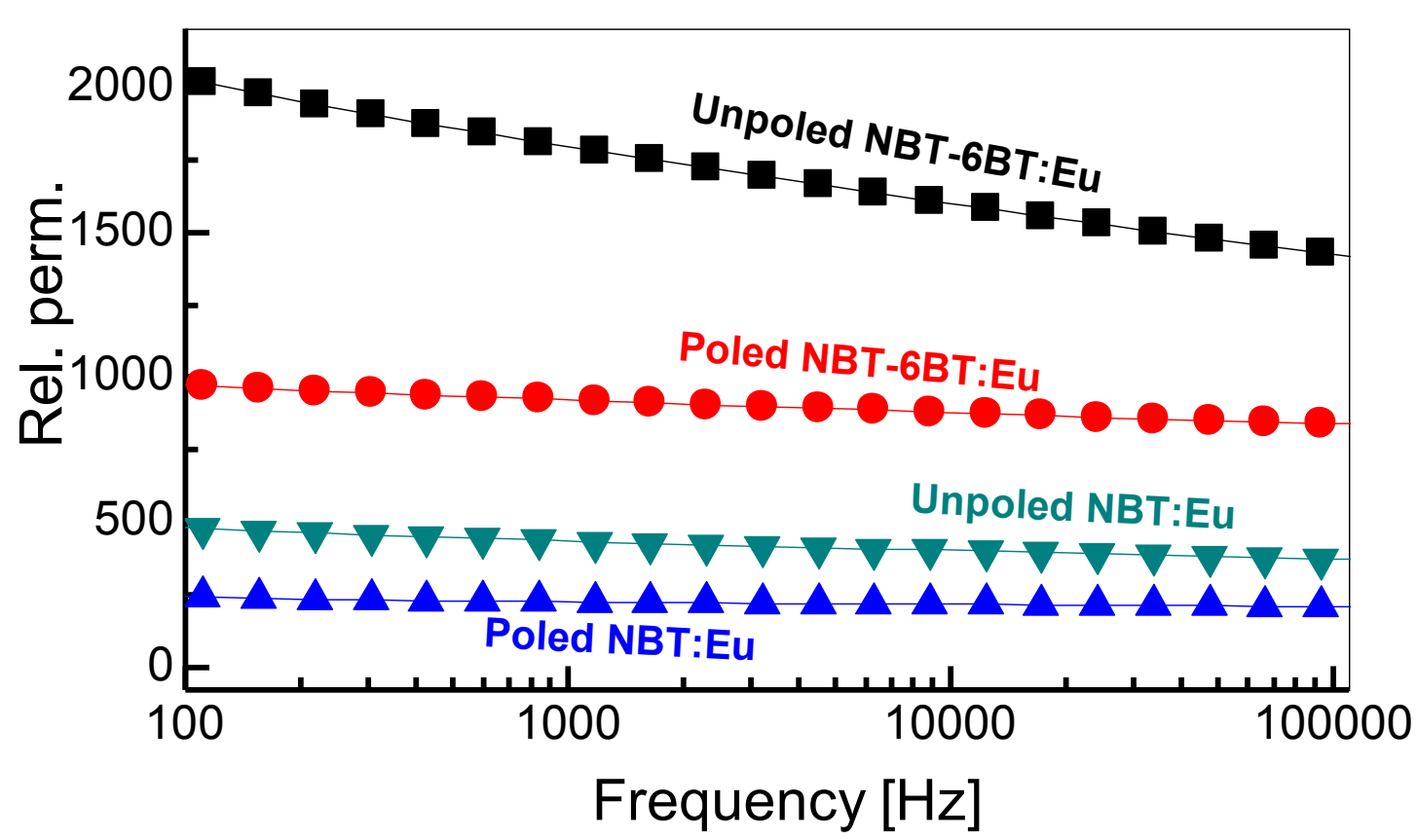

Fig. 4 Frequency dependence of the relative permittivities of NBT: Eu and NBT-6BT:Eu in the poled and unpoled state.

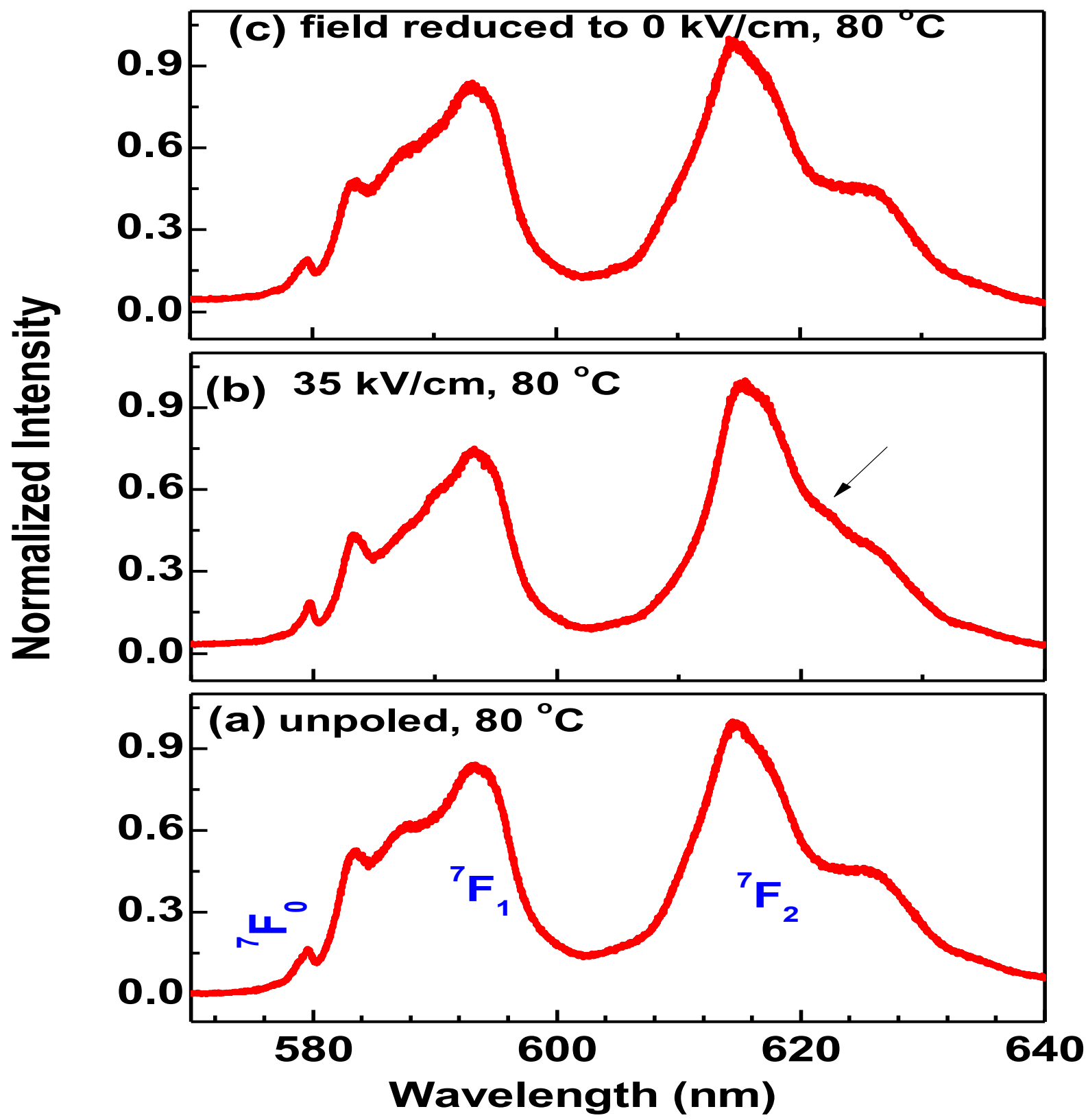

Fig. 5 $Eu^{+3}$ PL response with field in NBT-BT6:Eu. The specimen was kept at 80 C, close to the depolarization temperature (see S5). The PL spectrum was recorded at this temperature before the application of the field (a). PL spectrum at a field of 35 kV/cm at 80 $^0$C is shown in (b). The spectrtum after switching off the field is shown in (c). The arrow in (b) indicates the Stark new line due to electric field.